\documentclass{ieeeaccess}
\DeclareFontShape{T1}{formata}{m}{sl}{<->ssub*formata/m/it}{}
\DeclareFontShape{T1}{formata}{b}{it}{<->ssub*formata/b/n}{}
\usepackage{cite}
\usepackage{amsmath,amssymb,amsfonts}
\usepackage{array,booktabs}
\usepackage{graphicx}
\usepackage{url}
\usepackage{algorithm}
\usepackage{algpseudocode}

\usepackage{bm}
\makeatletter
\AtBeginDocument{\DeclareMathVersion{bold}
\SetSymbolFont{operators}{bold}{T1}{times}{b}{n}
\SetSymbolFont{NewLetters}{bold}{T1}{times}{b}{it}
\SetMathAlphabet{\mathrm}{bold}{T1}{times}{b}{n}
\SetMathAlphabet{\mathit}{bold}{T1}{times}{b}{it}
\SetMathAlphabet{\mathbf}{bold}{T1}{times}{b}{n}
\SetMathAlphabet{\mathtt}{bold}{OT1}{pcr}{b}{n}
\SetSymbolFont{symbols}{bold}{OMS}{cmsy}{b}{n}
\renewcommand\boldmath{\@nomath\boldmath\mathversion{bold}}}
\makeatother

\newcommand{\manuscripttablesetup}{\footnotesize}

\begin{document}

\history{Date of publication xxxx 00, 0000, date of current version xxxx 00, 0000.}
\doi{10.1109/ACCESS.xxxx.xxxxxxx}

\title{A Bayesian hierarchical model for meta-analysis}

\author{\uppercase{Jing Dai}\authorrefmark{1},
\uppercase{Sijie Xu}\authorrefmark{2}, AND
\uppercase{Shufei Ge}\authorrefmark{3}}

\address[1]{Department of International Education, Shanghai Pharmaceutical School, Shanghai, China}
\address[2]{School of Information Science and Technology, ShanghaiTech University, Shanghai, China}
\address[3]{Institute of Mathematical Sciences, ShanghaiTech University, Shanghai, China}

\tfootnote{Jing Dai and Sijie Xu contributed equally to this work.}

\markboth
{Dai \headeretal: Bayesian meta-analysis model}
{Dai \headeretal: Bayesian meta-analysis model}

\corresp{Corresponding author: Shufei Ge (e-mail: geshf@shanghaitech.edu.cn).}

\begin{abstract}
Meta-analysis is a key statistical tool for synthesizing clinical trial data to evaluate treatment effects, yet traditional methods such as fixed-effect and random-effects models often fail to handle heterogeneity, study-level covariates, or hierarchical structures effectively. To overcome these limitations, we developed a Bayesian hierarchical meta-analysis framework for robust parameter estimation on small samples and utilized analytical integration for efficient inference. Simulation studies indicated robust estimation of the proposed model. We applied it to the efficacy and safety profiles of Oxcarbazepine (OXC) and Carbamazepine (CBZ) in epilepsy treatment. Gibbs sampling and analytical integration produced nearly identical results across all three clinical datasets. The results indicated no significant difference between OXC and CBZ in complete seizure control, while OXC was significantly associated with a higher probability of at least a 50\% reduction in seizure frequency and a lower risk of side effects than CBZ. The code and relevant data used in this study are openly available on GitHub at \url{https://github.com/xsjk/HierarchicalMetaAnalysis}.
\end{abstract}


\titlepgskip=-21pt
\maketitle

\section{Introduction}
\PARstart{M}{eta-analysis} is a statistical technique that combines the results of multiple studies to identify patterns, assess consistency, and draw meaningful conclusions. It is widely applied in various fields, particularly in clinical trials \cite{antithrombotic2002collaborative,early2005effects,patel2022effect,cipriani2018comparative,parajuli2019effectiveness}, to assess the effectiveness of treatments, synthesize evidence, and support decision-making processes.
A large meta-analysis of 287 trials (involving 135,000 patients) \cite{antithrombotic2002collaborative} found that antiplatelet therapy, such as aspirin, lowered the risk of serious vascular events by about 25\%.
A meta-analysis of breast cancer treatment trials \cite{early2005effects} showed that chemotherapy and hormonal therapy greatly improved long-term survival.
A meta-analysis of seven COVID-19 trials \cite{patel2022effect} found that corticosteroids reduced the risk of death within 28 days. A network meta-analysis of 522 depression trials \cite{cipriani2018comparative} showed that all 21 antidepressants tested worked better than placebo. A meta-analysis of 16 trials \cite{parajuli2019effectiveness} found that pharmacist-led care for heart failure cut hospitalizations by around 28\%.

Carbamazepine (CBZ) and oxcarbazepine (OXC) are widely used antiepileptic drugs (AEDs) in clinical practice to reduce the frequency of seizures and improve the quality of life of patients suffering from epilepsy. These drugs can inhibit the flow of sodium ions and stabilize the neuronal cell membrane, thereby achieving the goal of suppressing seizures. They are commonly used to treat complex partial seizures, generalized tonic-clonic seizures, and other conditions.
CBZ is often recommended as the first-line antiepileptic drug treatment for patients with partial-onset seizures \cite{national2004diagnosis,french2004efficacy} due to its proven antiepileptic efficacy \cite{kosteljanetz1979carbamazepine,mattson1985comparison,mikkelsen1981clonazepam,ramsay1983double,rapeport1985factors,sillanpaa1979carbamazepine,simonsen1976comparative,troupin1977carbamazepine}. OXC was developed as a substitute for CBZ to improve safety. In vivo, OXC rapidly and completely converts to 10-OH-OXC \cite{kristensen1992pharmacokinetics,theisohn1982disposition}. In animal experiments, OXC and 10-OH-OXC showed antiepileptic efficacy similar to that of CBZ. Additionally, studies in healthy volunteers indicated that participants taking OXC experienced fewer side effects than those taking CBZ. However, it remains unclear whether OXC treatment is more effective and has fewer side effects than CBZ treatment for epilepsy in children, or the opposite.

The development and evaluation of AEDs require rigorous clinical trials to ensure their efficacy and safety. Meta-analysis plays a crucial role in synthesizing evidence from clinical trials and improving evidence-based recommendations. Traditional methods for meta-analysis, such as fixed-effect and random-effects models, have been widely used \cite{borenstein2010basic}. Fixed-effect models assume that all studies estimate the same effect size, providing precise estimates when heterogeneity is low \cite{zhai2024fixed}. However, this approach is not suitable for heterogeneous data \cite{lin2017alternative}. Random-effects models account for variability across studies but require accurate estimation of variance between studies, which can lead to imprecise results when the number of studies is small \cite{guolo2017random}. Both methods may struggle with incorporating complex study-level covariates or hierarchical data structures \cite{higgins2009re,fernandez2020application}. Hierarchical models can help address these challenges by accounting for heterogeneity between studies and populations, enabling enhanced understanding of treatment efficacy and facilitating personalized approaches to epilepsy management.

In this work, we propose an alternative method for conducting meta-analysis with a hierarchical model in a Bayesian framework and leverage inference techniques such as Markov Chain Monte Carlo (MCMC) \cite{metropolis1949monte} and analytical integration for parameter estimation. In addition, we apply our model to analyze the efficacy and safety of the antiepileptic drugs OXC and CBZ. The results indicate no significant difference between OXC and CBZ in complete seizure control, while OXC therapy has a significantly higher probability of at least a 50\% reduction in seizure frequency and a lower risk of side effects than CBZ therapy.

\section{Model} 
\label{model}

To compare the effectiveness of two treatments (treatment A and treatment B) for a given medical condition across independent studies, we first define the risk ratio (RR) for each study \(j=1,\dots,m\), where \(m\) is the total number of studies, as the ratio of the probability of the event occurring in group A to the probability of the event occurring in group B:
\begin{equation*}
    \text{RR}_j = \frac{\frac{{k_1}_j}{{n_1}_j}}{\frac{{k_2}_j}{{n_2}_j}},
    \quad j=1,\dots,m,
\end{equation*}

where \({k_1}_j\) and \({n_1}_j\) denote the number of events and total participants in group A of study \(j\), respectively; \({k_2}_j\) and \({n_2}_j\) denote the corresponding quantities for group B.

We further consider the logarithm of the risk ratio, defined as
\begin{equation*}
    y_j = \log\text{RR}_j = \log\left(\frac{{k_1}_j}{{n_1}_j}\right) - \log\left(\frac{{k_2}_j}{{n_2}_j}\right),
    \quad j=1,\dots,m.
\end{equation*}

By analyzing \(y_j\), we can assess whether treatment A is consistently more effective (or less effective) than treatment B for a given medical event.

To simplify the calculation, the variance of \(y_j\) can be approximated as follows \cite{deeks2010statistical}:

\begin{equation*}
   \hat{\sigma}_j^2 = \frac{1}{{k_1}_j} 
               - \frac{1}{{n_1}_j}
               + \frac{1}{{k_2}_j}
               - \frac{1}{{n_2}_j},
    \quad j=1,\dots,m.
\end{equation*}

To model the heterogeneity of different studies, we construct a hierarchical model as follows:
\begin{align*}
    y_j | \theta_j &\sim N(\theta_j, {\sigma}_j^2),\\
    \theta_j | \mu, \tau &\sim N(\mu, \tau^2),\\
    \mu &\sim N(\eta, \kappa^2),\\
    \tau^2 &\sim \Gamma^{-1}(\alpha, \beta).\\
    \text{for }j&=1,\dots,m.
\end{align*}

\begin{figure}[!t]
    \centering
    \includegraphics[width=0.9\columnwidth]{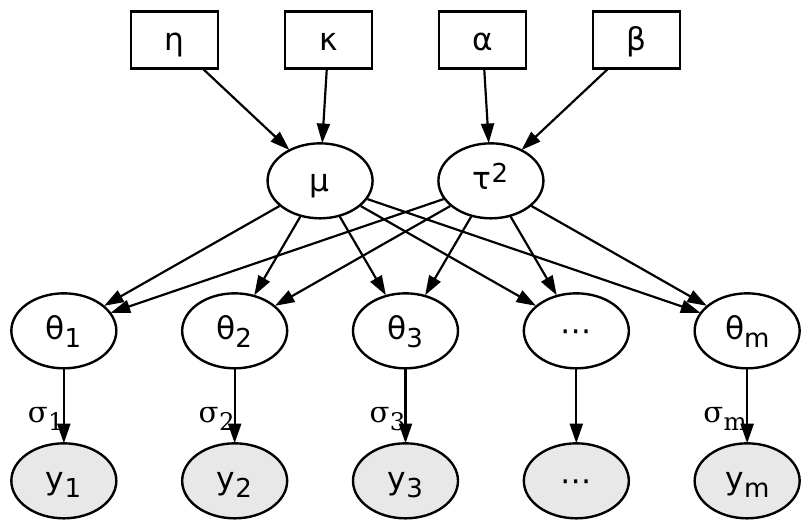}
    \caption{An illustration of the hierarchical model structure.}
    \label{fig:hierarchical-model}
\end{figure}

As illustrated in Figure \ref{fig:hierarchical-model}, the top-level parameters include \(\mu\), representing the global mean, and the global variance \(\tau^2\) modeled as an inverse-gamma distribution. The second-level parameters \(\theta_j\) are modeled as normal distributions influenced by \(\mu\) and \(\tau\), capturing group-level variability. The observed data \(y_j\) are generated from normal distributions centered around \(\theta_j\) with known variance \(\hat{\sigma}_j^2\), reflecting study-level observations. \(\eta\), \(\kappa\), \(\alpha\), and \(\beta\) are hyperparameters.

This approach provides a unified framework for comparison, accounting for study-specific sample sizes and event counts.

\section{Inference}
\subsection{Markov Chain Monte Carlo}
To estimate the parameters of the hierarchical model, we employ a Markov Chain Monte Carlo (MCMC) sampling method \cite{metropolis1949monte}. With the conjugate priors specified in Section \ref{model}, the derived posterior distributions of the model parameters are given below.

For \(\mu\mid\theta,\tau\),
\begin{equation}
    \mu\mid\theta,\tau
    \sim N\left(
        \mu_{\mu\mid\theta,\tau},
        \sigma^2_{\mu\mid\theta,\tau}
    \right),
    \label{eq:mu-gibbs}
\end{equation}
where
\begin{equation*}
    \mu_{\mu\mid\theta,\tau}
    =
    \frac{\frac{\eta}{\kappa^2}+\frac{\sum_{j=1}^m\theta_j}{\tau^2}}
         {\frac{1}{\kappa^2}+\frac{m}{\tau^2}},
    \qquad
    \sigma_{\mu\mid\theta,\tau}
    =
    \frac{1}{\sqrt{\frac{1}{\kappa^2}+\frac{m}{\tau^2}}}.
\end{equation*}

For \(\tau^2\mid\theta,\mu\),
\begin{equation}
    \tau^2\mid\theta,\mu
    \sim \Gamma^{-1}\left(
        \alpha+\frac{m}{2},
        \beta+\frac{1}{2}\sum_{j=1}^m(\theta_j-\mu)^2
    \right).
    \label{eq:tau-gibbs}
\end{equation}

For \(\theta_j\mid\mu,\tau,y_j\), \(j=1,\dots,m\),
\begin{equation}
    \theta_j\mid\mu,\tau,y_j
    \sim N\left(
        \mu_{\theta_j\mid\mu,\tau,y_j},
        \sigma^2_{\theta_j\mid\mu,\tau,y_j}
    \right),
    \label{eq:theta-gibbs}
\end{equation}
where
\begin{equation*}
    \mu_{\theta_j\mid\mu,\tau,y_j}
    =
    \frac{\frac{y_j}{\sigma_j^2}+\frac{\mu}{\tau^2}}
         {\frac{1}{\sigma_j^2}+\frac{1}{\tau^2}},
    \qquad
    \sigma_{\theta_j\mid\mu,\tau,y_j}
    =
    \frac{1}{\sqrt{\frac{1}{\sigma_j^2}+\frac{1}{\tau^2}}}.
\end{equation*}

Because the full conditional posterior of each parameter has an explicit formula, we implement Gibbs sampling to perform model inference, as shown in Algorithm \ref{gibbs}.

\begin{algorithm}[!t]
\footnotesize
\caption{Gibbs Sampling for the Hierarchical Model}
\label{gibbs}
\begin{algorithmic}[1]
    \State \textbf{Input:} \(y_1,\ldots,y_m\)
    \State \textbf{Output:} \(\{\theta_1^{(t)},\ldots,\theta_m^{(t)},\mu^{(t)},\tau^{2(t)}\}_{t=1}^{T}\)
    \State \textbf{Initialize:} \(\mu^{(0)}\), \(\tau^{2(0)}\), and \(\theta_j^{(0)}\), \(j=1,\ldots,m\)
    \For{\(t=1,\ldots,T\)}
        \State Sample \(\mu^{(t)}\sim p(\mu\mid\theta^{(t-1)},\tau^{(t-1)})\) using \eqref{eq:mu-gibbs}
        \State Sample \(\tau^{2(t)}\sim p(\tau^2\mid\theta^{(t-1)},\mu^{(t)})\) using \eqref{eq:tau-gibbs}
        \For{\(j=1,\ldots,m\)}
            \State Sample \(\theta_j^{(t)}\sim p(\theta_j\mid\mu^{(t)},\tau^{(t)},y_j)\) using \eqref{eq:theta-gibbs}
        \EndFor
    \EndFor
\end{algorithmic}
\end{algorithm}

\subsection{Analytical Method}

In this section, we outline an analytical method for more efficiently estimating the risk ratio at the group level (that is, \(\exp{\theta_j}\)).
The method takes advantage of the fact that \(\mu\), \(\theta_j\mid\mu\) and \(y_j\mid\theta_j\) follow normal distributions, from which we can find \(y_j\mid\mu,\tau\), \(\mu\mid\tau,y\), \(\theta_j\mid\mu,\tau,y\) and \(\theta_j\mid\tau,y\), which are all Gaussian, enabling the marginal posterior distributions of \(\tau\), \(\mu\), and \(\theta_j\) to be computed without the need for complex multidimensional integrals. Instead, an explicit formula without integral can be derived for the marginal posterior of \(\tau\), and \(\mu\) and \(\theta_j\) can be expressed through single integrals over \(\tau\). This approach allows for efficient numerical computation of the expectations, variances, and credible intervals of the parameters, with the final estimates of the risk ratio \(\mathrm{e}^{\mu}\mid y\) and \(\mathrm{e}^{\theta_j}\mid y\) for \(j=1,\dots,m\), directly available from these posterior distributions.

\subsubsection{Marginal posterior of \(\tau\)}

From the distribution of \(y_j\mid\theta_j\) and \(\theta_j\mid\mu,\tau\) for \(j=1,\dots,m\), we can find that \(y_j\mid\mu,\tau\) follows a normal distribution,
\begin{equation*}
    y_j | \mu, \tau \sim N\big(\mu, \hat{\sigma}_j^2 + \tau^2\big) ,
    \quad j=1,\dots,m.
\end{equation*}
Then, we can derive the marginal posterior of \(\tau\).
\begin{align*}
    f(\tau | y)
        &\propto f(\tau) \int_{-\infty}^\infty f(\mu)
            \prod_{j=1}^m f(y_j | \mu, \tau)\,\mathrm{d}\mu \\
        &\propto f(\tau)
            \frac{\exp\!\left[-\frac{1}{2}\left\{\nu_2(\tau)
                -\frac{\nu_1(\tau)^2}{\nu_0(\tau)}\right\}\right]}
            {\sqrt{\nu_0(\tau)\Pi(\tau)}}.
\end{align*}
For the inverse-gamma prior on \(\tau^2\), the density in this expression is the induced density
\begin{equation*}
    f(\tau)=2\tau f_{\tau^2}(\tau^2)
    =\frac{2\beta^\alpha}{\Gamma(\alpha)}
      \tau^{-2\alpha-1}\exp\left(-\frac{\beta}{\tau^2}\right),
    \quad \tau>0.
\end{equation*}
The analytical expression also applies when \(f(\tau)\) is replaced by another prior density. Define
\begin{align*}
    \nu_k(\tau) &= \sum_{j=1}^m \frac{y_j^k}{\hat{\sigma}_j^2+\tau^2}+\frac{\eta^k}{\kappa^2}, \quad k=0,1,2,\\
    \Pi(\tau) &= \prod_{j=1}^m \left(\hat{\sigma}_j^2+\tau^2\right).
\end{align*}
Here, \(0^0\) is defined as 1.

\subsubsection{Marginal posterior of \(\mu\)}

With the distribution of \(y\mid\mu,\tau\) and the prior distribution of \(\mu\), we can find that \(\mu\mid\tau,y\) follows a normal distribution,
\begin{equation*}
    \mu | \tau, y \sim  N(\mu_{\mu | \tau, y}, \sigma^2_{\mu | \tau, y} ),
\end{equation*}
where
\begin{equation*}
    \mu_{\mu | \tau, y} = \frac{\nu_1(\tau)}{\nu_0(\tau)},\quad
    \sigma_{\mu | \tau, y}= \frac{1}{\sqrt{\nu_0(\tau)}}.
\end{equation*}
Using the marginal posterior of \(\tau\), we can derive the marginal posterior of \(\mu\), including its moments and equal-tailed 95\% credible interval (CI).
\begin{align*}
    f(\mu | y) 
        &= \int_0^\infty f(\mu | \tau, y) f(\tau | y)\,\mathrm{d}\tau, \\
    F(\mu | y) 
        &= \int_0^\infty F(\mu | \tau, y) f(\tau | y)\,\mathrm{d}\tau, \\
   \mathbb{E}[\mu | y] 
        &= \int_0^\infty f(\tau|y)\mu_{\mu | \tau, y}\,\mathrm{d}\tau, \\
    \mathrm{Var}(\mu|y) 
        &= \int_0^\infty f(\tau|y) \left(\mu_{\mu | \tau, y}^2 + \sigma^2_{\mu | \tau, y}\right)
        \mathrm{d}\tau - \mathbb{E}[\mu | y]^2, \\
    \mathrm{CI}_{0.95}(\mu | y) &= \left[F_{\mu|y}^{-1}(0.025), F_{\mu|y}^{-1}(0.975)\right].
\end{align*}

\subsubsection{Marginal posterior of \(\theta_j\)}

With the distribution of \(\theta_j\mid\mu,\tau,y\) and the distribution of \(\mu\mid\tau,y\), we can find that \(\theta_j\mid\tau,y\) for \(j=1,\dots,m\) follows a normal distribution,
\begin{equation*}
    \theta_j | \tau, y \sim N\left(\mu_{\theta_j | \tau, y}, \sigma^2_{\theta_j | \tau, y} \right),
\end{equation*}
where
\begin{equation*}
    \mu_{\theta_j | \tau, y} = \frac{\omega_{j1}(\tau)}{\omega_{j0}(\tau)},\quad 
    \sigma_{\theta_j | \tau, y} = \frac{1}{\sqrt{\omega_{j0}(\tau)}},
\end{equation*}
\begin{align*}
        \omega_{jk}(\tau) &= \frac{y_j^k}{\hat{\sigma}_j^2}
            + \frac{\tilde{\nu}_{jk}(\tau)}{
                \tau^2\tilde{\nu}_{j0}(\tau) + 1
            }, \\
        \tilde{\nu}_{jk}(\tau) &= \nu_k(\tau) - \frac{y_j^k}{\hat{\sigma}_j^2 + \tau^2},
           \quad \text{for }k=0,1.\\
\end{align*}
Using the marginal posterior of \(\tau\), we can derive the marginal posterior of \(\theta_j\) for \(j=1,\dots,m\):
\begin{align*}
    f(\theta_j | y) 
        &= \int_0^\infty f(\theta_j | \tau, y)f(\tau | y)\,\mathrm{d}\tau,\\
    F(\theta_j|y)
        &= \int_0^\infty F(\theta_j | \tau, y)f(\tau | y)\,\mathrm{d}\tau,\\
   \mathbb{E}[\theta_j | y] 
        &= \int_0^\infty f(\tau|y)\mu_{\theta_j | \tau, y}\,\mathrm{d}\tau,\\
    \mathrm{Var}(\theta_j|y) 
        &= \int_0^\infty f(\tau|y)
            \left(\mu^2_{\theta_j | \tau, y}+\sigma^2_{\theta_j | \tau, y}\right)
            \,\mathrm{d}\tau \\
        &\quad - \mathbb{E}[\theta_j | y]^2,\\
    \mathrm{CI}_{0.95}(\theta_j | y) &= \left[F_{\theta_j|y}^{-1}(0.025), F_{\theta_j|y}^{-1}(0.975)\right].
\end{align*}

\section{Simulation Study}

To evaluate the performance of the proposed Bayesian hierarchical meta-analysis model, we conducted three scenarios with varying degrees of parameter uncertainty. Each simulation was repeated 12,800 times to measure the stability of the estimates using the coverage rates of the 95\% credible intervals, bias, and root mean squared error (RMSE). In each replication, posterior means obtained from the posterior distributions derived in the preceding section were used as point estimates, and equal-tailed 95\% credible intervals were used to calculate coverage; bias and RMSE compare the point estimates with the corresponding true values. The standard deviation parameters \(\sigma_j\) for \(j=1,\dots,9\) were set as constants in the simulations as \([0.233,\allowbreak 0.165,\allowbreak 0.169,\allowbreak 0.058,\allowbreak 0.835,\allowbreak 0.218,\allowbreak 0.157,\allowbreak 0.285,\allowbreak 0.199]\), corresponding to nine hypothetical studies in the Application section to mimic the real data. Priors were set as \(\mu\sim N(0,1)\) and \(\tau^2\sim\Gamma^{-1}(0.001,0.001)\).

\subsection{Simulations}

Since the risk ratio \(\text{RR}=e^\mu\) and study-specific risk ratios \(\text{RR}_j=e^{\theta_j}\) for \(j=1,\dots,m\) are monotonic transformations of \(\mu\) and \(\theta_j\), their coverage rates are equivalent to those of the original parameters. Thus, we omit separate reporting of \(RR\) and \(RR_j\) metrics in the tables below.

\subsubsection{Simulation 1: (Fixed \(\mu\), \(\tau\))}

For each trial \(k\), both \(\mu=0.14\) and \(\tau=0.1\) were fixed. Group-level effects \(\theta_{k,j}\) were generated from \(\mathcal{N}(\mu,\tau^2)\), and data \(y_{k,j}\) from \(\mathcal{N}(\theta_{k,j},\sigma_j^2)\). Table \ref{tab:simulation1} reports the coverage rate, bias, and RMSE for all parameters in Simulation 1.

\begin{table}[!htbp]
\centering
\caption{Simulated results for \(\mu=0.14\) and \(\tau=0.1\)}
\label{tab:simulation1}
\manuscripttablesetup
\begin{tabular}{@{}lccc@{}}
\toprule
\textbf{Variable} & \textbf{Coverage Rate} & \textbf{Bias} & \textbf{RMSE} \\
\midrule
\(\theta_1\) & 0.965 & 0.001 & 0.115 \\
\(\theta_2\) & 0.955 & -0.001 & 0.105 \\
\(\theta_3\) & 0.958 & 0.000 & 0.106 \\
\(\theta_4\) & 0.950 & 0.001 & 0.054 \\
\(\theta_5\) & 0.979 & -0.002 & 0.123 \\
\(\theta_6\) & 0.962 & 0.000 & 0.114 \\
\(\theta_7\) & 0.956 & 0.000 & 0.103 \\
\(\theta_8\) & 0.968 & -0.001 & 0.119 \\
\(\theta_9\) & 0.959 & 0.001 & 0.111 \\
\(\mu\) & 0.968 & -0.001 & 0.068 \\
\(\tau\) & 0.991 & 0.025 & 0.054 \\
\bottomrule
\end{tabular}
\end{table}

\subsubsection{Simulation 2: (Sampled \(\mu\), Fixed \(\tau\))}

For each trial \(k\), \(\mu_k\) was sampled from \(\mathcal{N}(0,1)\) while \(\tau=0.1\) remained fixed. Group-level effects \(\theta_{k,j}\) were generated from \(\mathcal{N}(\mu_k,\tau^2)\), and observed data \(y_{k,j}\) were generated from \(\mathcal{N}(\theta_{k,j},\sigma_j^2)\). Table \ref{tab:simulation2} reports the coverage rate, bias, and RMSE for all parameters in Simulation 2.

\begin{table}[!htbp]
\centering
\caption{Simulated results for \(\mu\sim\mathcal{N}(0, 1)\) and \(\tau=0.1\).}
\label{tab:simulation2}
\manuscripttablesetup
\begin{tabular}{@{}lccc@{}}
\toprule
\textbf{Variable} & \textbf{Coverage Rate} & \textbf{Bias} & \textbf{RMSE} \\
\midrule
\(\theta_1\) & 0.961 & 0.000 & 0.117 \\
\(\theta_2\) & 0.958 & 0.000 & 0.104 \\
\(\theta_3\) & 0.959 & -0.001 & 0.105 \\
\(\theta_4\) & 0.952 & 0.001 & 0.054 \\
\(\theta_5\) & 0.978 & -0.001 & 0.124 \\
\(\theta_6\) & 0.962 & 0.001 & 0.114 \\
\(\theta_7\) & 0.954 & 0.002 & 0.103 \\
\(\theta_8\) & 0.964 & 0.001 & 0.120 \\
\(\theta_9\) & 0.958 & 0.000 & 0.112 \\
\(\mu\) & 0.971 & 0.000 & 0.068 \\
\(\tau\) & 0.990 & 0.025 & 0.054 \\
\bottomrule
\end{tabular}
\end{table}

\subsubsection{Simulation 3: (Sampled \(\mu\) and \(\tau\))}

For each trial \(k\), \(\mu_k\) was generated from \(\mathcal{N}(0,1)\) and \(\tau_k\) was generated from \(\text{Uniform}(0,1)\). Group-level effects \(\theta_{k,j}\) were generated from \(\mathcal{N}(\mu_k,\tau_k^2)\), and observed data \(y_{k,j}\) were generated from \(\mathcal{N}(\theta_{k,j},\sigma_j^2)\). Table \ref{tab:simulation3} reports the coverage rate, bias, and RMSE for all parameters in Simulation 3.

\begin{table}[!htbp]
\centering
\caption{Simulation results for \(\mu\sim\mathcal{N}(0, 1)\) and \(\tau\sim \text{Uniform}(0, 1)\).}
\label{tab:simulation3}
\manuscripttablesetup
\begin{tabular}{@{}lccc@{}}
\toprule
\textbf{Variable} & \textbf{Coverage Rate} & \textbf{Bias} & \textbf{RMSE} \\
\midrule
\(\theta_1\) & 0.944 & 0.000 & 0.203 \\
\(\theta_2\) & 0.944 & -0.001 & 0.152 \\
\(\theta_3\) & 0.945 & 0.001 & 0.156 \\
\(\theta_4\) & 0.949 & 0.000 & 0.057 \\
\(\theta_5\) & 0.940 & -0.008 & 0.470 \\
\(\theta_6\) & 0.944 & 0.005 & 0.194 \\
\(\theta_7\) & 0.948 & -0.002 & 0.145 \\
\(\theta_8\) & 0.942 & 0.000 & 0.238 \\
\(\theta_9\) & 0.945 & 0.002 & 0.179 \\
\(\mu\) & 0.944 & -0.001 & 0.206 \\
\(\tau\) & 0.928 & 0.032 & 0.190 \\
\bottomrule
\end{tabular}
\end{table}

\subsection{Result Interpretation}

The method demonstrated strong performance for \(\mu\) and \(\theta_j\), with coverage rates near 95\% and RMSE proportional to \(\sigma_j\) (Tables \ref{tab:simulation1}--\ref{tab:simulation3}). The \(\tau\) estimates had coverage rates ranging from 92.8\% to 99.1\%, positive biases ranging from 0.025 to 0.032, and RMSE values ranging from 0.054 to 0.190.

\section{Application}\label{sec:application}

To compare the efficacy and safety between OXC and CBZ, different electronic databases were searched for relevant studies with a cut-off date of 15 August 2023, including the Cochrane Controlled Trials Register in the Cochrane Library (Issue 8, 2023), MEDLINE, EMBASE.com, China National Knowledge Infrastructure (CNKI), and Wanfang Data. The search terms used were ``oxcarbazepine, carbamazepine, epilepsy'', and the search was limited to studies involving pediatric patients and clinical trials. Reference lists of relevant reviews and retrieved articles were also checked for further identification. The published abstracts and proceedings were checked to follow up on ongoing clinical trials. If necessary, the manufacturers of OXC and CBZ were contacted.

We set the following inclusion criteria: (i) children with all forms of onset seizures diagnosed based on the 1981 International League Against Epilepsy (ILAE) classification \cite{ZHSJ200103026}; (ii) OXC or CBZ monotherapy; (iii) double-blinded, single-blinded, or unblinded trials were included; (iv) adequately randomized and quasi-randomized trials were included; (v) articles published in English or Chinese were included; (vi) if the outcome of one clinical trial was published multiple times, we included the latest one. Trials that met any of the following conditions were excluded: (i) non-randomized and non-controlled trials; (ii) trials including patients with intractable epilepsy who were awaiting epilepsy surgery or already had surgery; (iii) animal trials; (iv) trials including patients over the age of 18.

A total of 1191 clinical trials and reports were reviewed, and nine randomized controlled trials (RCTs) \cite{donati2007cognitive,chen2012clinical,chen2013clinical,zhang2015effect,ZWVJ201610036,ZJZH201610022,TDYX201608010,YBQJ201602046,HYYX202006014} met our selection criteria and were chosen for the meta-analysis. Outcomes from each trial were extracted for subsequent analysis. These nine studies provided a comprehensive dataset, as shown in Table \ref{tab:raw-data}, which was further subdivided into three distinct datasets, each focusing on one of the following outcomes: complete control, 50\% reduction, and side effects.

The Jadad composite scale, which assesses the description of randomization, blinding, and withdrawals in reports \cite{altman2001concealing}, was used to score the quality of included studies \cite{jadad1996assessing}. The quality scale ranges from 0 to 5 points, with a score of 1 or less indicating a low-quality report, a score of 2 indicating a medium-quality report, and a score of at least 3 indicating a high-quality report \cite{moher1998does}. As shown in Table \ref{tab:quality-scores}, all nine included RCTs \cite{donati2007cognitive,chen2012clinical,chen2013clinical,zhang2015effect,ZWVJ201610036,ZJZH201610022,TDYX201608010,YBQJ201602046,HYYX202006014} had medium- or high-quality Jadad scores.

\section{Results}

\begin{table}[!htbp]
\centering
\caption{Comparison of treatment outcomes and side effects between Oxcarbazepine and Carbamazepine across studies in children with epilepsy. Each cell reports four values: complete seizure control events / events with at least a 50\% reduction in seizure frequency / side-effect events / total participants.}
\label{tab:raw-data}
\manuscripttablesetup
\begin{tabular}{@{}lcc@{}}
\toprule
\textbf{Study} & \textbf{Oxcarbazepine} & \textbf{Carbamazepine} \\
\midrule
\cite{donati2007cognitive}  & 32 /  - / 17 / 55 & 13 /  - /  8 / 28 \\
\cite{ZJZH201610022}        & 29 / 43 /  7 / 45 & 27 / 39 / 26 / 45 \\
\cite{HYYX202006014}        & 26 / 35 /  7 / 40 & 25 / 33 / 19 / 40 \\
\cite{zhang2015effect}      & 80 / 87 /  - / 88 & 62 / 71 /  - / 72 \\
\cite{ZWVJ201610036}        &  4 / 34 /  4 / 36 &  2 / 29 / 11 / 39 \\
\cite{TDYX201608010}        & 21 / 32 /  - / 40 & 20 / 27 /  - / 40 \\
\cite{YBQJ201602046}        & 41 / 59 /  4 / 61 & 30 / 50 / 23 / 60 \\
\cite{chen2012clinical}     & 16 / 25 /  5 / 32 & 12 / 21 /  8 / 30 \\
\cite{chen2013clinical}     & 30 / 43 / 11 / 60 & 25 / 38 / 23 / 58 \\
\bottomrule
\end{tabular}
\end{table}

\begin{table*}[!t]
\centering
\caption{Jadad quality scores of the RCTs included in the meta-analysis.}
\label{tab:quality-scores}
\manuscripttablesetup
\begin{tabular}{@{}llclcc@{}}
\toprule
\textbf{Ref.} & \textbf{Randomization} & \textbf{Blinding} &
\textbf{Withdrawals and dropouts} & \textbf{Jadad Score} & \textbf{Quality} \\
\midrule
\cite{donati2007cognitive} & Randomization mentioned, but method not specified & Not performed & Clearly reported & 2 & Medium \\
\cite{chen2012clinical} & Closed-envelope lottery & Not performed & Clearly reported & 3 & High \\
\cite{chen2013clinical} & Closed-envelope lottery & Not performed & Clearly reported & 3 & High \\
\cite{zhang2015effect} & Random number & Not performed & Clearly reported & 3 & High \\
\cite{ZWVJ201610036} & Randomization mentioned, but method not specified & Not performed & Clearly reported & 2 & Medium \\
\cite{ZJZH201610022} & Random number table & Not performed & Clearly reported & 3 & High \\
\cite{TDYX201608010} & Randomization mentioned, but method not specified & Not performed & Clearly reported & 2 & Medium \\
\cite{YBQJ201602046} & Randomization mentioned, but method not specified & Not performed & Clearly reported & 2 & Medium \\
\cite{HYYX202006014} & Randomization mentioned, but method not specified & Not performed & Clearly reported & 2 & Medium \\
\bottomrule
\end{tabular}
\end{table*}

We adopt the following hyperparameters to assign relatively non-informative priors to both \(\mu\) and \(\tau\): \(\alpha=\beta=0.001\), \(\eta=0\), and \(\kappa=1\). This model is applied to the three separate datasets derived from the nine studies, using two distinct inference methods: the sampling method (with 100,000 steps of Monte Carlo sampling) and the analytical method (using numerical integration and solving numerical equations). For each of these three datasets, both methods yield nearly identical results. The sample means, standard deviations, and credible interval endpoints obtained from the sampling method align closely with those from the analytical method, with maximum discrepancies of 0.003, 0.017, and 0.023, respectively.

\subsection{Complete Controls Comparison}

In this section, the number of patients with clinical efficacy was taken as the primary outcome. Clinical efficacy was defined as complete control of seizures during the study. Clinical efficacy was reported in all nine RCTs \cite{donati2007cognitive,chen2012clinical,chen2013clinical,zhang2015effect,ZWVJ201610036,ZJZH201610022,TDYX201608010,YBQJ201602046,HYYX202006014}, involving 869 patients with epilepsy. For these nine RCTs, OXC and CBZ were compared. As shown in Table \ref{tab:complete-control}, the overall \(\mathrm{RR}=1.119\) (\(95\%~\mathrm{CI}=[0.990,~1.275]\)), indicating no significant difference in clinical efficacy between OXC and CBZ therapies.

\begin{table}[!htbp]
\centering
\caption{The nine trials comparing the effect of OXC and CBZ on complete control of seizure in children with epilepsy.}
\label{tab:complete-control}
\manuscripttablesetup
\scriptsize
\setlength{\tabcolsep}{1.2pt}
\begin{tabular}{@{}lcccccccc@{}}
\toprule
\textbf{Study or Subgroup} & \(j\) & \multicolumn{2}{c}{\textbf{OXC}} & \multicolumn{2}{c}{\textbf{CBZ}} & \multicolumn{3}{c}{\textbf{RR}} \\
 & & \textbf{Events} & \textbf{Total} & \textbf{Events} & \textbf{Total} & \textbf{Mean} & \textbf{Std} & \textbf{95\% CI} \\

\midrule
\cite{donati2007cognitive}  & 1 & 32 & 55 & 13 & 28  & 1.136 & 0.114 & [0.945,~ 1.403] \\
\cite{ZJZH201610022}        & 2 & 29 & 45 & 27 & 45  & 1.111 & 0.096 & [0.931,~ 1.318] \\
\cite{HYYX202006014}        & 3 & 26 & 40 & 25 & 40  & 1.105 & 0.096 & [0.921,~ 1.308] \\
\cite{zhang2015effect}      & 4 & 80 & 88 & 62 & 72  & 1.082 & 0.056 & [0.975,~ 1.193] \\
\cite{ZWVJ201610036}        & 5 &  4 & 36 &  2 & 39  & 1.133 & 0.135 & [0.916,~ 1.443] \\
\cite{TDYX201608010}        & 6 & 21 & 40 & 20 & 40  & 1.111 & 0.105 & [0.916,~ 1.340] \\
\cite{YBQJ201602046}        & 7 & 41 & 61 & 30 & 60  & 1.162 & 0.112 & [0.987,~ 1.431] \\
\cite{chen2012clinical}     & 8 & 16 & 32 & 12 & 30  & 1.132 & 0.118 & [0.934,~ 1.407] \\
\cite{chen2013clinical}     & 9 & 30 & 60 & 25 & 58  & 1.126 & 0.106 & [0.941,~ 1.366] \\
\midrule
Overall                     &   &    &    &    &     & 1.119 & 0.072 & [0.990,~ 1.275] \\
\bottomrule
\end{tabular}
\end{table}

\subsection{Half Reductions Comparison}

The 50\% reduction in seizure frequency and frequency of side effects were considered secondary outcomes. Eight RCTs involving 786 patients reported the 50\% reduction in seizure frequency \cite{chen2012clinical,chen2013clinical,zhang2015effect,ZWVJ201610036,ZJZH201610022,TDYX201608010,YBQJ201602046,HYYX202006014}. As shown in Table \ref{tab:half-reduction}, the overall \(\mathrm{RR}=1.096\) (\(95\%~\mathrm{CI}=[1.012,~1.200]\)), indicating that OXC therapy had a significantly higher probability of reduced seizure frequency than CBZ therapy.

\begin{table}[!htbp]
\centering
\caption{The eight trials comparing the effect of OXC and CBZ on the 50\% reduction in seizure frequency in children with epilepsy.}
\label{tab:half-reduction}
\manuscripttablesetup
\scriptsize
\setlength{\tabcolsep}{1.2pt}
\begin{tabular}{@{}lcccccccc@{}}
\toprule
\textbf{Study or Subgroup} & \(j\) & \multicolumn{2}{c}{\textbf{Oxcarbazepine}} & \multicolumn{2}{c}{\textbf{Carbamazepine}} & \multicolumn{3}{c}{\textbf{Risk Ratio}} \\
 & & \textbf{Events} & \textbf{Total} & \textbf{Events} & \textbf{Total} & \textbf{Mean} & \textbf{Std} & \textbf{95\% CI} \\

\midrule
\cite{ZJZH201610022}    & 1 & 43 & 45 & 39 & 45  & 1.097 & 0.057 & [0.993,~ 1.217] \\
\cite{HYYX202006014}    & 2 & 35 & 40 & 33 & 40  & 1.082 & 0.067 & [0.956,~ 1.223] \\
\cite{zhang2015effect}  & 3 & 87 & 88 & 71 & 72  & 1.010 & 0.018 & [0.974,~ 1.046] \\
\cite{ZWVJ201610036}    & 4 & 34 & 36 & 29 & 39  & 1.151 & 0.085 & [1.013,~ 1.345] \\
\cite{TDYX201608010}    & 5 & 32 & 40 & 27 & 40  & 1.116 & 0.085 & [0.973,~ 1.309] \\
\cite{YBQJ201602046}    & 6 & 59 & 61 & 50 & 60  & 1.128 & 0.059 & [1.023,~ 1.254] \\
\cite{chen2012clinical} & 7 & 25 & 32 & 21 & 30  & 1.100 & 0.084 & [0.951,~ 1.288] \\
\cite{chen2013clinical} & 8 & 43 & 60 & 38 & 58  & 1.095 & 0.078 & [0.954,~ 1.265] \\
\midrule
Overall                 &   &    &    &    &     & 1.096 & 0.048 & [1.012,~ 1.200] \\
\bottomrule
\end{tabular}
\end{table}

\subsection{Side Effects Comparison}

Seven RCTs involving 629 patients with epilepsy reported side effects for OXC and CBZ \cite{donati2007cognitive,chen2012clinical,chen2013clinical,ZWVJ201610036,ZJZH201610022,YBQJ201602046,HYYX202006014}. As shown in Table \ref{tab:side-effects}, the overall \(\mathrm{RR}=0.456\) (\(95\%~\mathrm{CI}=[0.291,~0.692]\)), indicating that OXC therapy had a significantly lower risk of side effects than CBZ therapy.

\begin{table}[!htbp]
\centering
\caption{The seven trials comparing the safety of OXC and CBZ on the frequency of side effects in children with epilepsy.}
\label{tab:side-effects}
\manuscripttablesetup
\scriptsize
\setlength{\tabcolsep}{1.2pt}
\begin{tabular}{@{}lcccccccc@{}}
\toprule
\textbf{Study or Subgroup} & \(j\) & \multicolumn{2}{c}{\textbf{Oxcarbazepine}} & \multicolumn{2}{c}{\textbf{Carbamazepine}} & \multicolumn{3}{c}{\textbf{Risk Ratio}} \\
 & & \textbf{Events} & \textbf{Total} & \textbf{Events} & \textbf{Total} & \textbf{Mean} & \textbf{Std} & \textbf{95\% CI} \\

\midrule
\cite{donati2007cognitive}  & 1 & 17 & 55 &  8 & 28  & 0.672 & 0.270 & [0.358,~ 1.371] \\
\cite{ZJZH201610022}        & 2 &  7 & 45 & 26 & 45  & 0.382 & 0.105 & [0.191,~ 0.599] \\
\cite{HYYX202006014}        & 3 &  7 & 40 & 19 & 40  & 0.429 & 0.113 & [0.234,~ 0.683] \\
\cite{ZWVJ201610036}        & 4 &  4 & 36 & 11 & 39  & 0.451 & 0.146 & [0.221,~ 0.800] \\
\cite{YBQJ201602046}        & 5 &  4 & 61 & 23 & 60  & 0.361 & 0.122 & [0.139,~ 0.603] \\
\cite{chen2012clinical}     & 6 &  5 & 32 &  8 & 30  & 0.505 & 0.177 & [0.269,~ 0.961] \\
\cite{chen2013clinical}     & 7 & 11 & 60 & 23 & 58  & 0.464 & 0.112 & [0.284,~ 0.729] \\
\midrule
Overall                     &   &    &    &    &     & 0.456 & 0.106 & [0.291,~ 0.692] \\
\bottomrule
\end{tabular}
\end{table}

\section{Discussion}

This study uses two methods, MCMC and numerical integration, to analyze the marginal posterior distributions in a meta-analysis of antiepileptic drugs. MCMC provides flexibility in handling complex models, while numerical integration offers computational efficiency but may struggle with high-dimensional parameter spaces.

Epilepsy remains a common disease that is usually treated with medication. In recent decades, many new AEDs have been developed to reduce the frequency of seizures. However, CBZ remains the first choice. Moreover, OXC is frequently chosen as a substitute for CBZ because it is considered to be effective and to have fewer side effects. In this meta-analysis study, after rigorous quality control and application of the selection criteria, nine RCTs were included to compare the safety and effectiveness of OXC and CBZ monotherapies for seizures in children. A total of 869 patients were included in these RCTs conducted in Europe and China. Our findings suggest that there was no significant difference between OXC and CBZ in complete seizure control, while OXC had a significantly higher probability of at least a 50\% reduction in seizure frequency and a lower risk of side effects than CBZ in children. More high-quality trials from other countries are necessary to provide stronger evidence for choosing between CBZ and OXC in the treatment of seizures in children. Future work will explore hybrid inference methods that combine MCMC and numerical integration for hierarchical models and extend the analysis to other AEDs or broader patient populations to validate these findings.

\bibliographystyle{IEEEtran}
\bibliography{references}

\begin{IEEEbiographynophoto}{Jing Dai}
Jing Dai received a MSc degree in Clinical Pharmacy and a B.S. degree in Pharmacy from West China School of Pharmacy, Sichuan University, China. She also received a MSc degree in Polymer Material Science and Engineering from the School of Materials, the University of Manchester, UK. She joined the Shanghai Pharmaceutical School in July 2015.
\end{IEEEbiographynophoto}

\begin{IEEEbiographynophoto}{Sijie Xu}
received the B.S. degree in computer science from ShanghaiTech University. He is currently pursuing the M.S. degree in computer science with the School of Information Science and Technology, ShanghaiTech University. His research interests include data science, parallel computing, and computational mathematics.
\end{IEEEbiographynophoto}

\begin{IEEEbiographynophoto}{Shufei Ge} received her Ph.D. in Statistics from Simon Fraser University in 2020, and joined ShanghaiTech University as an Assistant Professor in the same year. Her research interests lie at the intersection of Bayesian statistics, statistical machine learning, bioinformatics, and statistical genetics. In particular, she focuses on developing scalable statistical methodologies to address pressing challenges in computational biology, including neuroimaging genetics and complex disease prediction. Her work aims to bridge advanced statistical theory with practical applications in modern biomedical research.
\end{IEEEbiographynophoto}

\EOD

\end{document}